# Black Phosphorus p-MOSFETs with High Transconductance and Nearly Ideal Subthreshold Slope

Nazila Haratipour, Matthew C. Robbins, and Steven J. Koester

*Abstract*—We report record performance for black phosphorus p-MOSFETs. The devices have locally patterned back gates and 20-nm-thick $HfO_2$ gate dielectrics. Devices with effective gate length, $L_{eff}$ = 1.0 μm display extrinsic transonductance, $g_m$ of 101 μS/μm at a drain-to-source bias voltage, $V_{ds}$ = −3 V. Temperature-dependent analysis also shows that the subthreshold slope, *SS*, is nearly ideal, with a minimum value of *SS* = 66 mV/decade at room temperature and $V_{ds}$ = −0.1 V. Furthermore, devices with 7-nm $HfO_2$ dielectrics and $L_{eff}$ = 0.3 μm displayed $g_m$ as high as 204 μS/μm at $V_{ds}$ = −1.5 V.

## I. INTRODUCTION

TWO-dimensional (2D) semiconductors are of great interest due to their potential to realize extreme-scaled metal-oxide field-effect transistors (MOSFETs). The primary 2D semiconductors that have been studied recently are the transition-metal dichalcogenides ($MX_2$), particularly $MoS_2$, $WSe_2$ and $MoTe_2$ [1]-[4]. However, these semiconductors have relatively high effective mass and theoretical predictions suggest that $MX_2$ MOSFETs may be better suited for applications rather than high-performance logic [5]. Recently, several studies on black phosphorus (BP) MOSFETs have been reported [6]-[10]. BP is the most stable and least reactive isotope of phosphorus under standard conditions. This material has a layered crystal structure like that of $MX_2$ and can exist as a single, atomically-thin sheet (phosphorene). However unlike $MX_2$ semiconductors, BP is predicted to have a much lighter effective mass than $MX_2$ ($0.08m_0$ along one of the in-plane directions) and the mass is expected to be highly anisotropic with crystal orientation [11]. In addition, its band bap is expected to increase with decreasing layer thickness between about 0.3 eV and 1.0−1.5 eV [10],[12]. While 2D MOSFETs with thin dielectrics have been fabricated using graphene [13] and $MoS_2$ [14], to date, despite having high drive current, the BP MOSFETs reported in the literature have shown relatively low transconductance, $g_m$, and shallow subthreshold slope. In this work, we report the first BP MOSFETs using local back gates and thin high-K dielectrics and these devices display extrinsic $g_m$ over 200 μS/μm with similar devices displaying nearly-ideal linear-region subthreshold slope.

This work was partially supported by C-SPIN, one of the six SRC STARnet Centers, sponsored by MARCO and DARPA, and also by the NSF under grant No. ECCS-1102278. This work also utilized the University of Minnesota Nanofabrication and Characterization Facilities, which receive partial support from the National Science Foundation.

N. Haratipour, M. C. Robbins, and S. J. Koester are with the Department of Electrical and Computer Engineering, University of Minnesota, 200 Union St. SE, Minneapolis, MN 55455 (e-mail: skoester@umn.edu).

## II. DEVICE FABRICATION

The locally-backgated BP MOSFET fabrication sequence was similar to the process utilized previously for graphene-based varactors [15]. First, electron beam lithography (EBL) was used to pattern PMMA openings on a $Si/SiO_2$ wafer with $SiO_2$ thickness of 300 nm. Next, the $SiO_2$ was recessed by 50 nm using a combination of dry and wet etching. Ti (10 nm) and Pd (40 nm) were then deposited and lifted off such that the gate metal was roughly planar with the surrounding $SiO_2$. Then, 20 nm of $HfO_2$ was deposited at 300 °C using atomic layer deposition (ALD). BP flakes were mechanically exfoliated from bulk crystals and then transferred to a PDMS stamp on a glass slide. Thin flakes on the PDMS stamp were identified under an optical microscope, and aligned and transferred onto the embedded gates using a micro-positioner. Atomic-force microscopy of similar flakes transferred onto bare $Si/SiO_2$ substrates showed the BP thickness was roughly ~ 12 nm. Finally, EBL was utilized to pattern source and drain contact openings, followed by deposition and lift-off of Ti (5 nm) / Au (100 nm). Optical micrographs of the devices after BP alignment and source/drain metallization are shown in Fig. 1.

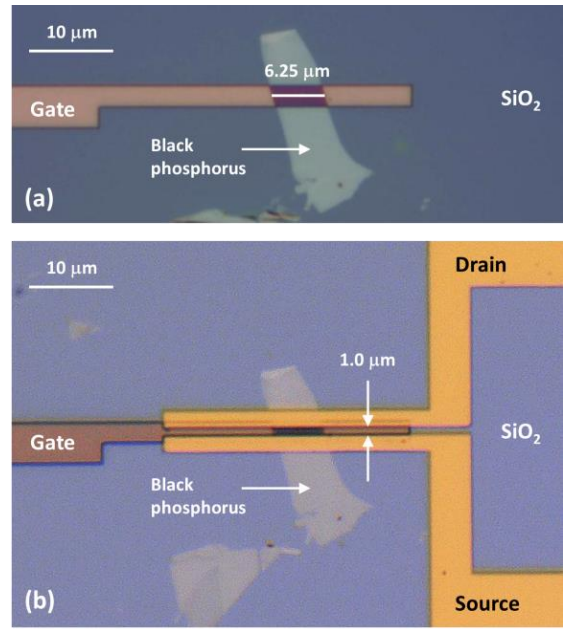

FIG. 1. Optical micrographs of locally-backgated BP MOSFETs (a) after flake transfer and (b) completed device after source/drain pad metallization. The gate width is 6.25 μm and the gate length is 1.0 μm. The gate metal is Pd with 20 nm $HfO_2$ gate dielectric.

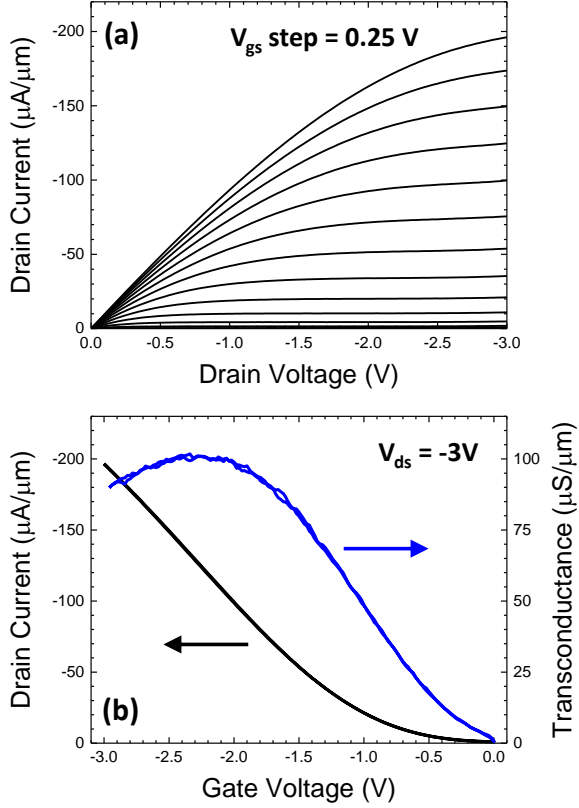

FIG. 2. (a) Drain current, $I_d$, vs. drain-to-source voltage, $V_{ds}$, at room temperature of a BP p-MOSFET with $L_{eff} = 1$ μm and HfO$_2$ gate dielectric thickness of 20 nm. The maximum drain current is 196 μA/μm at $V_{gs} = V_{ds} = -3$ V. (b) $I_d$ and transconductance, $g_m$, vs. $V_{gs}$ characteristic at room temperature of same device in (a) at $V_{ds} = -3$ V, for both gate voltage sweep directions. The peak $g_m$ of 101 μS/μm occurs at $V_{gs} = -2.4$ V.

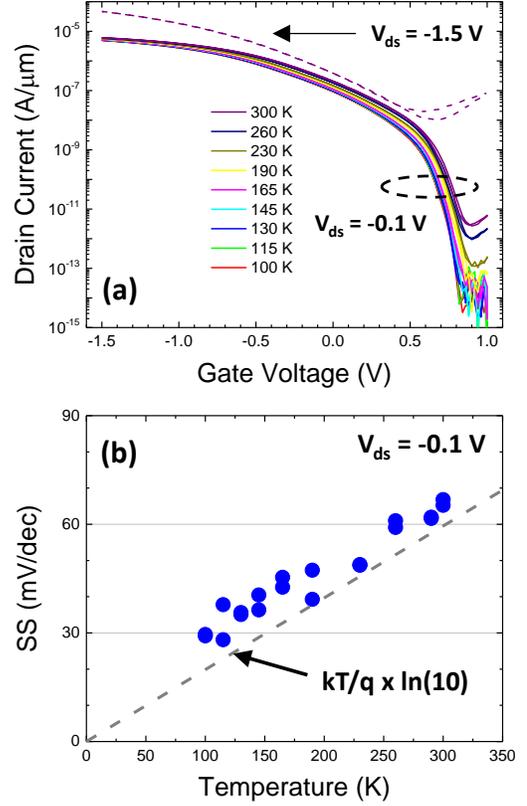

FIG 3. (a) Semilog plot of $I_d$ vs. $V_{gs}$ at 300 K of a BP p-MOSFET for $V_{ds} = -0.1$ V (purple – solid) and -1.5 V (purple – dashed) for both gate voltage sweep directions. The plot also shows the $I_d$ vs. $V_{gs}$ characteristics at $V_{ds} = -0.1$ V from $T = 100$ K to 300 K. (b) Minimum subthreshold slope (SS) vs. temperature at $V_{ds} = -0.1$ V for up and down sweep directions. The average SS at $T = 300$ K is 66 mV/decade.

All of the devices were transferred into a vacuum chamber immediately after fabrication. The devices analyzed in this work had channel width of 6.25 μm and effective gate length, $L_{eff}$, of 1.0 μm, where $L_{eff}$ was defined as the source-to-drain metallization spacing. While this design is not practical for high-speed devices due to the large overlap capacitance, it is suitable to explore the DC characteristics of the devices.

## III. RESULTS

All device characterization was performed using an Agilent B1500A semiconductor parameter analyzer. The devices were probed using a Lakeshore CPX-VF cryogenic probe station at temperatures, $T$, ranging from 100 to 300 K. All measurements were performed in vacuum. The room-temperature output characteristics of the device are shown in Fig. 2(a). The device shows nearly-ideal long-channel MOSFET behavior with excellent current saturation. The drive current is 196 μA/μm at gate-to-source and drain-to-source voltages of $V_{gs} = -3$ V and $V_{ds} = -3$ V, respectively. The room-temperature transfer characteristics are shown in Fig. 2(b) for both forward and reverse sweep directions. The devices show virtually no hysteresis and are consistent with the $I_d$-$V_{ds}$ data in Fig. 2(a). The threshold voltage extracted from standard long-channel analysis is roughly +0.4 V, indicating the devices are slightly depletion-mode. Also shown in Fig. 2(b) is a plot of extrinsic transconductance, $g_m$, vs. $V_{gs}$ at various $V_{ds}$ values. The peak extrinsic $g_m$ was 101 μS/μm at $V_{ds} = -3$V, and $V_{gs} = -2.4$ V.

The room-temperature linear ($V_{ds} = -0.1$ V) and saturation ($V_{ds} = -1.5$ V) subthreshold curves are shown in Fig. 3(a), along with the temperature dependence of the linear characteristic between $T = 100$ and 300 K. It should be noted that both sweep directions (increasing and decreasing $V_{gs}$) are shown in this plot, further highlighting the very low hysteresis in our devices. Extraction of the hole mobility, $\mu_h$, from the linear trasconductance showed values of $\mu_h = 65 \pm 7$ cm$^2$/Vs ($59 \pm 7$ cm$^2$/Vs) at $T = 300$ (100 K), where the equivalent oxide thickness (EOT) was assumed to be $4.5 \pm 0.5$ nm. This is a reasonable assumption based upon our previous data on graphene transistors using a similar dielectric thickness. The reason for the low $\mu_h$ is unclear, though we note that the crystal orientation of the BP is not known. Our observations for the value and temperature-dependence of the mobility are in contrast to recent values for BP on SiO$_2$ [10]. Possible reasons for this discrepancy could be the different dielectrics, as well as degradation of the BP due to moisture absorption during fabrication [16].

The minimum value of SS at $V_{ds} = -0.1$ V was extracted at each temperature and the results are plotted in Fig. 3(b) for both sweep directions. The results show nearly ideal behavior over all temperatures with an average value of 66 mV/decade at room temperature. This result shows that the interface between the BP and the HfO$_2$ has very low interface state density, and highlights the advantage of the inverted-gate geometry.

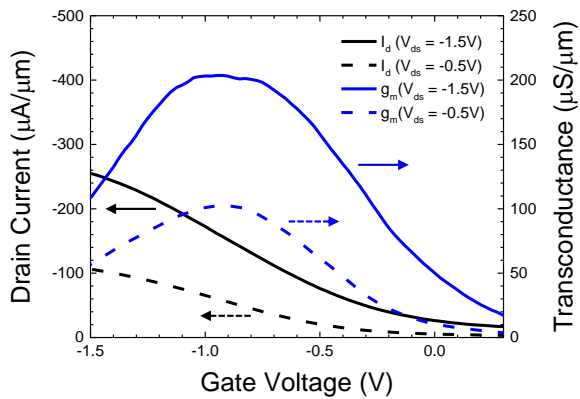

FIG. 4. $I_d$ and $g_m$, vs. $V_{gs}$ characteristic at room temperature a BP p-MOSFET with $L_{eff}$ = 0.3 μm and HfO$_2$ gate dielectric thickness of 7 nm at room temperature. The peak extrinsic transconductance at $V_{ds}$ = −1.5 V (−0.5 V) is 204 μS/μm (102 μS/μm).

Finally, we have also fabricated BP p-MOSFETs with HfO$_2$ thickness of 7 nm and $L_{eff}$ of 0.3 μm and the results for these devices are shown in Fig. 4. The devices show higher performance than the 1.0-μm gate-length devices with $g_m$ values up to 204 μS/μm (102 μS/μm) at $V_{ds}$ = −1.5 V (−0.5 V). However, the on-to-off current ratio was degraded compared to the longer-channel devices and at high drain bias, the devices did not completely pinch off. Despite the higher $g_m$, this value is lower than expectations given their shorter gate length and thinner dielectric compared to the devices in Fig. 2. This could possibly be due to series resistance effects or degraded mobility compared to the longer gate-length devices. Therefore, improvements in the processing conditions may be needed to minimize degradation of the BP during device processing.

## IV. CONCLUSION

In conclusion, we have fabricated BP p-MOSFETs with locally-patterned back gate electrodes and thin high-K dielectrics and demonstrated devices with extremely-high transconductance and nearly-ideal subthreshold slope. These results provide strong evidence that black phosphorus is a promising material for future high-performance CMOS.